\documentclass[runningheads]{llncs}
\usepackage{geometry}
\usepackage{cite}
\usepackage{amsmath,amssymb,amsfonts}
\usepackage{mathtools}
\usepackage{graphicx}
\usepackage{textcomp}
\usepackage{xcolor}
\def\BibTeX{{\rm B\kern-.05em{\sc i\kern-.025em b}\kern-.08em
    T\kern-.1667em\lower.7ex\hbox{E}\kern-.125emX}}

\usepackage{algorithm}
\usepackage{algorithmicx}
\usepackage{algpseudocode}

\usepackage{booktabs}
\usepackage{tabularx}
\newcolumntype{b}{X}
\newcolumntype{t}{>{\hsize=.15\hsize}X}
\newcolumntype{s}{>{\hsize=.25\hsize}X}
\newcolumntype{q}{>{\hsize=.35\hsize}X}
\newcolumntype{k}{>{\hsize=.8\hsize}X}
\newcolumntype{y}{>{\hsize=.15\hsize\centering\arraybackslash}X}

\usepackage{adjustbox}
\usepackage{tikz}
\newcommand*\circled[4]{\tikz[baseline=(char.base)]{
    \node[shape=circle, fill=#2, draw=#3, text=#4, inner sep=1pt] (char) {#1};}}
\usetikzlibrary{chains,positioning,arrows.meta,positioning,quotes,shapes}
\usetikzlibrary{shapes.multipart,positioning,fit,calc,decorations.pathmorphing}
\usepackage{hyperref}
\usepackage{url}

\usepackage{enumitem}

\newlist{questions}{enumerate}{2}
\setlist[questions,1]{label=Q\arabic*.,ref=Q\arabic*}
\setlist[questions,2]{label=(\alph*),ref=\thequestionsi(\alph*)}

\begin{document}
\title{Finding Privacy-relevant Source Code\thanks{Accepted by \textit{The 2nd International Workshop on Mining Software Repositories Applications for Privacy and Security} (MSR4P\&S 2024).}}
\titlerunning{Finding Privacy-relevant Source Code}
%
\author{Feiyang Tang\and Bjarte M. \O stvold }
\authorrunning{F. Tang and B.M. \O stvold}
%
\institute{Norwegian Computing Center\\N-0314 Oslo, Norway \\
\email{\{feiyang,bjarte\}@nr.no}}
\maketitle              
\begin{abstract}
Privacy code review is a critical process that enables developers and legal experts to ensure compliance with data protection regulations. However, the task is challenging due to resource constraints. To address this, we introduce the concept of \emph{privacy-relevant methods} --- specific methods in code that are directly involved in the processing of personal data. We then present an automated approach to assist in code review by identifying and categorizing these privacy-relevant methods in source code.

Using static analysis, we identify a set of methods based on their occurrences in 50 commonly used libraries. We then rank these methods according to their frequency of invocation with actual personal data in the top 30 GitHub applications. The highest-ranked methods are the ones we designate as privacy-relevant in practice.
For our evaluation, we examined 100 open-source applications and found that our approach identifies fewer than 5\% of the methods as privacy-relevant for personal data processing. This reduces the time required for code reviews. Case studies on Signal Desktop and Cal.com further validate the effectiveness of our approach in aiding code reviewers to produce enhanced reports that facilitate compliance with privacy regulations.

\end{abstract}

\section{Introduction}
In the realm of software development, privacy code reviews have become indispensable, especially with the advent of stringent data protection regulations like the General Data Protection Regulation (GDPR). Unlike security code reviews, which focus on existing security flaws or vulnerabilities, privacy code reviews are concerned with the ethical and lawful handling of personal data. Although there may be overlaps, such as in access control, the primary objectives of these two types of reviews are distinct: security reviews aim to prevent unauthorized access, while privacy reviews aim for compliance with data protection principles.

\begin{figure*}[h]
    \centering
    \includegraphics[width=\textwidth]{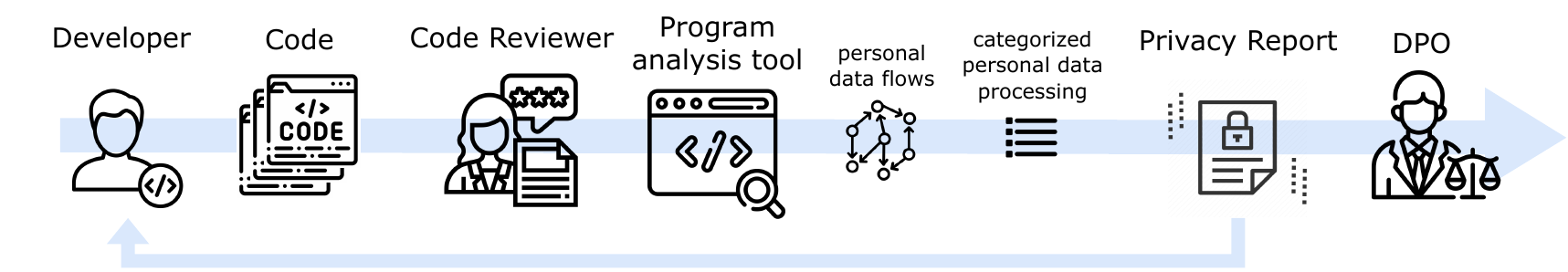}
    \caption{Privacy code review process}
    \label{fig:pcr}
\end{figure*}

Privacy code reviews involve a systematic process where source code is inspected to trace the flow of personal data. Equipped with program analysis tools, reviewers categorize these flows and detail how personal data is processed. This analysis serves as a comprehensive guide for compliance checks and aids Data Protection Officers (DPOs) in fulfilling their responsibilities. The process is illustrated in Figure~\ref{fig:pcr}. However, the challenge arises from the complexity and sheer volume of modern codebases, making it difficult to identify instances where personal data is processed.

Recent studies~\cite{10.1145/2906388.2906392,van2022detecting} have examined tools for identifying personal data, but less focus has been placed on data that is dynamically changing or in active use. While categorizations exist for personal data itself, taxonomies of the processing code are lacking. Developing a understanding of the diverse ways data can be handled would illuminate processing activities and facilitate compliance reporting like records of processing activities (ROPA) and data protection impact assessments (DPIA). Since reviewing entire codebases is time-consuming, targeting reports to highlight the most relevant aspects could better serve reviewers and streamline the compliance process. The goal should be providing clarity on key data handling activities without getting lost in an elaborate labeling framework.

In light of these challenges, we propose an automated approach to enhance the efficiency and effectiveness of privacy code reviews. Our approach focuses on identifying \emph{privacy-relevant methods} --- specifically, Java methods or JavaScript functions commonly found in popular libraries --- that are involved in the processing of personal data. By doing so, we can pinpoint instances in real-world applications where these privacy-relevant methods are invoked to handle personal data.

This paper addresses the following research questions:
\begin{enumerate}
    \item How to identify privacy-relevant methods in commonly used libraries that potentially process personal data? \label{rq1}
    \item How to categorize such privacy-relevant methods based on their actual usage in real-world applications? \label{rq2}
\end{enumerate}

To answer these questions, we make the following contributions:
\begin{enumerate}
    \item We present a novel static analysis technique specifically designed to identify methods in source code that are involved in the processing of personal data. (Section~\ref{Sec:identify})
    \item We develop a set of labels for categorizing personal data and the methods that process them, thereby providing a structured approach to understanding how personal data is processed in code. (Sections~\ref{Sec:taxonomy} and \ref{Sec:IdentifyingPersonalData})
    \item We apply our approach to a set of popular open-source applications. Through this, we rank privacy-relevant methods based on their frequency of occurrence, thereby identifying those that are most critical for privacy considerations. (Section~\ref{Sec:experiment})
    \item We provide insights to code reviewers by highlighting frequently used methods relevant to privacy, based on our large-scale study and specific case studies. This approach streamlines the review process, enabling a more focused and efficient identification of potential privacy risks. (Section~\ref{Sec:finding})
\end{enumerate}

Our evaluation of 100 open-source applications indicates that our approach identifies fewer than \textbf{5\%} of methods involved in personal data processing as privacy-relevant methods.This enables reviewers to focus only on the identified relevant code, thereby expediting privacy code reviews.

\section{Background}\label{Sec:background}
Code review, originally aimed at ensuring software quality by identifying bugs and performance issues~\cite{thongtanunam2020review}, has expanded to address security vulnerabilities and, more recently, privacy concerns under data protection laws like the GDPR. Privacy-focused reviews add the complexity of ensuring personal data is handled lawfully and ethically, a challenging task due to the often ambiguous nature of data protection guidelines~\cite{Tang_2023}.

Static analysis tools are pivotal in code reviews, aiding in the identification of data flows, security risks, and compliance issues. The effectiveness of a review is measured by its ability to pinpoint critical problems and offer actionable solutions. Privacy code reviews, however, struggle with identifying personal data due to unclear definitions and varied contexts, increasing reliance on these tools despite their limitations in recognizing diverse personal data types~\cite{icissp23}.

These reviews also play a key role in creating essential compliance documents like Records of Processing Activities (ROPA) and Data Protection Impact Assessments (DPIA). The proposed automated approach in this paper focuses on improving the efficiency and accuracy of privacy code reviews, specifically in categorizing personal data processing in large-scale code projects.

\section{Privacy-Relevant Methods} \label{Sec:definition}

To streamline the process of privacy code review, we introduce the concept of \textit{privacy-relevant methods}. These are specific methods that play a direct role in the processing of personal data. Such methods can be part of standard libraries or third-party libraries, making them critical focal points for personal data processing in software applications.

Native libraries are foundational because they offer the only pathways to device resources like files and networks. Consequently, any operation involving data storage or transfer must go through these native methods. Native privacy-relevant methods are those found in standard libraries of programming languages like JavaScript and Java. These methods act as the origins (sources) for all personal data entered by users via devices. They are also the exclusive methods that directly transmit this data to other devices or services. We categorize these native methods into domains such as \emph{I/O, Database, Network, Security}, following the guidelines of existing research~\cite{10.1145/3549035.3561185}. We identify these methods through a systematic manual review that includes an examination of documentation, source code, and actual usage patterns.

To facilitate the identification and categorization of native privacy-relevant methods, we conducted an in-depth analysis of key modules like \texttt{java.io}, \texttt{java.security}, and \texttt{java.util} for Java, and their equivalents in JavaScript. This analysis helps us compile a complete set of native privacy-relevant methods, denoted as $\mathit{Native}$, that are involved in personal data processing.

\section{Identifying API Privacy-relevant Methods} \label{Sec:identify}

Native privacy-relevant methods form the basis for identifying what we refer to as API privacy-relevant methods. These are methods found in third-party libraries and frameworks that are likely to process personal data by calling upon native privacy-relevant methods. 
Understanding the relationship between API and native methods is crucial for a complete review of how personal data is processed in a codebase.
The identification process is iterative and takes into account the dependencies between libraries and codebases, as depicted in Fig.~\ref{fig:flowchart}. The goal is to assemble a list of API privacy-relevant methods that have the potential to handle personal data. Understanding the relationship and dependency hierarchy among these libraries is essential for accomplishing this task.

\usetikzlibrary {shapes.multipart,positioning,fit,calc,decorations.pathmorphing}
\begin{figure}[ht]
\centering
\begin{tikzpicture}[decoration=snake,node distance=15mm and 25mm]
\node (mn) [fill=red!20,draw,double,rounded corners] {Native Privacy-relevant Methods};
\node (ma) [fill=red!20,draw,double,rounded corners,below=of mn.west,anchor=west] {API Privacy-relevant Methods};
\node [draw=none,rounded corners=0.3cm,fill opacity=0.1,fill=red,inner sep=10pt,fit={(mn)(ma)},label={[red]above:Privacy-relevant Methods}] {};
\node (mp) [fill=blue!20,draw,double,rounded corners, right=of mn, xshift=-5mm, yshift=-6mm] {Application Code};
\draw[->,thick,densely dashed](mp.west) -- (mn.east);
\draw[->,thick,densely dashed](mp.west) -- (ma.east) node[midway,below,yshift=-.5mm,xshift=2mm] {invoke};
\draw[->,thick,densely dashed](ma) -- (mn) node[midway,right] {invoke};
\end{tikzpicture}
\caption{The relationships between privacy-relevant methods and application code}
\label{fig:flowchart}
\end{figure}
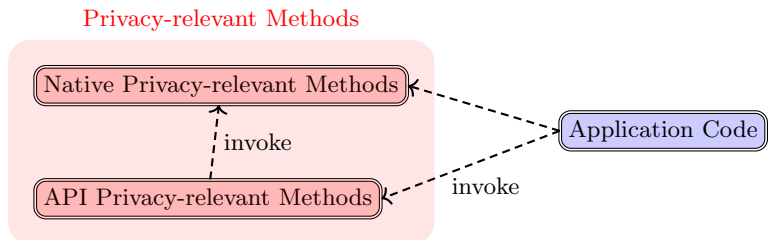

\subsection{Dependency Sorting and Identification of Privacy-relevant Methods} \label{Sec:dependency}
To manage library dependencies, we focus on import statements within each library's source code. We organize the libraries in a sequence such that each library is evaluated only after all its dependencies have been assessed. This ensures a logical and efficient evaluation process.

For the identification of API privacy-relevant methods, we define a set denoted as $\mathit{API}$. This set includes methods from our organized list of libraries that invoke native privacy-relevant methods at some point during their execution. These methods are significant as they interact with native methods, either directly or through a chain of calls, making them critical for privacy code review.

\section{Labels for Personal Data Processing}\label{Sec:taxonomy}

Compliance with data protection regulations like GDPR necessitates a nuanced understanding of how personal data is processed within code. While GDPR outlines various processing activities such as collection, recording, and organization, the four native privacy-relevant method categories~\cite{10.1145/3549035.3561185} we previously discussed (\textit{I/O}, \textit{security}, \textit{database}, and \textit{network}) lack the granularity needed for comprehensive understanding. For instance, the \textit{security} category encompasses both authentication and encryption, warranting a more detailed labeling system.

After analyzing top labels from Maven and NPM that pertain to personal data processing, we identified 20 labels that closely align with both GDPR's definitions and our native privacy-relevant method categories. This shows how libraries handle data processing in different ways. For example, \textit{OAuth} combines \textit{network} and \textit{security} functionalities, while \textit{Object-Relational Mapping (ORM)} bridges \textit{database} and \textit{I/O operations}. These overlaps underscore the necessity for a detailed set of labels tailored for privacy reviews. We present these labels and their alignment with GDPR requirements in Table~\ref{table:taxonomy-gdpr}.

\begin{table}[ht]
\centering
\caption{Alignment of the labels with GDPR requirements}
\setlength{\tabcolsep}{10pt} 
\begin{tabularx}{\columnwidth}{kq}
\toprule
\textbf{Category} & \textbf{GDPR Alignment} \\
\midrule
\textbf{Identity and Access Management (IAM)}:\newline managing users' identities and regulating their access to resources. It is based on the libraries that perform authentication and access control. & Article 32: Robust measures, including authentication. \\
\midrule
\textbf{Data Encryption and Cryptography (DEC)}:\newline cryptographic operations enhancing the security and privacy of personal data. DEC specifically targets data encryption. & Article 32: Data pseudonymization and encryption. \\
\midrule
\textbf{Data Storage, Management, and Deletion (DSMD)}:\newline handling the storage, retrieval, and deletion of personal data. DSMD extends the ``\textit{database}'' category to incorporate data deletion methods, focusing on the lifecycle management of personal data in storage systems. & Article 5(1)(e): Data retention only as long as necessary. \\
\midrule
\textbf{Data Processing and Transformation (DPT)}:\newline carry out transformations or processing on personal data, including anonymization, aggregation, and other forms of data manipulation. DPT also includes Object/Relational Mapping methods that facilitate the conversion between incompatible type systems in object-oriented programming languages and databases. & Article 30: Mandatory record of processing activities under responsibility. \\
\midrule
\textbf{Network Communication (NC)}:\newline methods that send or receive personal data over a network, focusing on personal data transmission. & Article 44: Controlled data transfer. \\
\midrule
\textbf{Logging and Monitoring (LM)}:\newline methods that handle the recording, monitoring, and retrieval of log entries that may contain personal data. It emphasizes the tracking and auditing activities that involve personal data. & Artical 5(1)(c): Principle of data minimization. Article 5(1)(e): Data retention only as long as necessary.
\\
\bottomrule
\end{tabularx}
\label{table:taxonomy-gdpr}
\end{table}

These labels serve a dual purpose: they categorize methods involved in data processing activities like collection, storage, and encryption, and they map these activities to GDPR compliance requirements. This streamlined mapping simplifies the task of identifying code sections that need to comply with legal standards. In our later approach, we use these labels to prioritize privacy-relevant methods, enabling a focused review on areas critical for data protection.

\section{Process of Identifying Personal Data} \label{Sec:IdentifyingPersonalData}

Before delving into the approach, it is crucial to differentiate between personal data and personally identifiable information (PII). While both are subsets of information that relate to an individual, PII is a category of data that directly identifies a person. Examples include account information, contact details, personal IDs, and national IDs. 
Not all the 10 categories of personal data we consider below fall under PII. The exposure of PII is especially concerning as it could lead to personal or psychological harm, such as identity theft.

Our primary aim is to identify the flow of personal data within a codebase, focusing on its cruicial implications for privacy. To achieve this, we use a pattern-matching technique inspired by Tang et al.~\cite{tang2023helping}. This technique effectively identifies data from 10 categories, including \textit{Account}, \textit{Contact}, \textit{Personal ID}, \textit{Location}, and \textit{National ID}. We employ Semgrep, a tool tailored for pattern matching in code, to facilitate this process. Semgrep's rules are specifically designed for Java and JavaScript languages.


\subsection{Static Analysis for Personal Data Identification} 

The initial phase of our approach involves using static analysis to locate code fragments that contain personal data. We use Semgrep for this task, given its efficiency and flexibility in analyzing large codebases. We rely on Semgrep's support for multiple languages and its capabilities for local data flow analysis.

\subsection{Defining Sources of Personal Data} 

In the context of our analysis, sources refer to instances where personal data appears. We identify personal data in two ways: 1) as literal text present in the source code, and 2) as variables, based on their name identifiers. Our identification rules are designed to support Java, JavaScript, and TypeScript but can be extended to other languages that Semgrep supports.

\subsection{Rule Crafting for Identification}

To pinpoint literal personal data, we use regular expression (regex) matching. This comes into play, for example, when identifying the format of national ID numbers. For variable sources, we maintain a default list of identifiers that correspond to the 10 categories of personal data. These identifiers help us formulate Semgrep rules. To reduce false positives, we impose specific conditions on these regex rules. For instance, to capture all human names in the code, we use a regex pattern that accommodates variations like first, last, and full names: \texttt{(?i).(?:first|given|full|last|sur(?!geon))} \texttt{[s/(;)|,=!>]name)}.

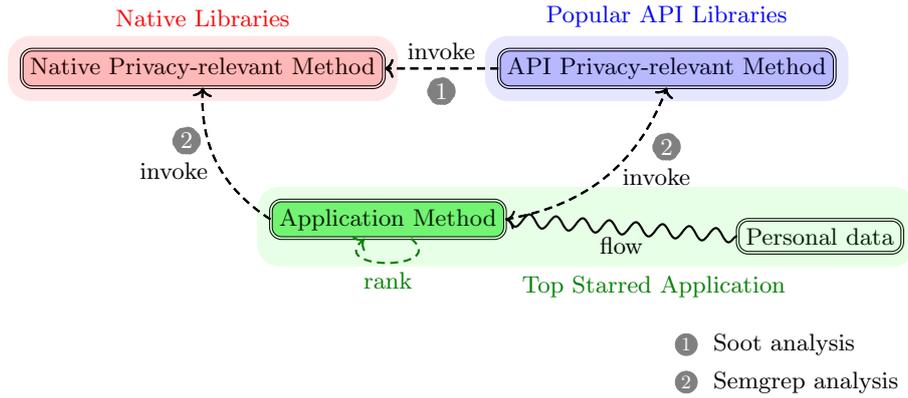
\begin{figure}[h!]
\centering
\begin{tikzpicture}[decoration=snake,node distance=20mm and 30mm]
\node (mn2) [fill=red!20,draw,double,rounded corners,anchor=west,align=left] {Native Privacy-relevant Method};
\node (native) [draw=none,rounded corners=0.3cm,fill opacity=0.1,fill=red,inner sep=5pt,fit={(mn2)},label={[red]above: Native Libraries}] {};
\node (ma2) [fill=blue!20,draw,double,rounded corners,anchor=west,right=of mn2.east,xshift=-1.5cm,align=left] {API Privacy-relevant Method};
\node (api) [draw=none,rounded corners=0.3cm,fill opacity=0.1,fill=blue,inner sep=5pt,fit={(ma2)},label={[blue]above:Popular API Libraries}] {};
\node (mapp2) [fill=black!10!green!50,draw,double,rounded corners,below=of mn2.east,anchor=west,align=left,xshift=-1.5cm] {Application Method};
\node (pd) [draw,double,rounded corners,below=of ma2.east,xshift=-.2cm,yshift=0cm,align=left] {Personal data};
\node (app) [draw=none,rounded corners=0.3cm,fill opacity=0.1,fill=green,inner sep=5pt,fit={(mapp2)(pd)},label={[black!50!green,xshift=-1.5cm]below right: Top Starred Application}] {};
\draw [->,decorate,thick](pd.west) -- (mapp2.east) node[midway,below,xshift=0cm] {flow};
\draw [->,thick,densely dashed](ma2.west) -- (mn2.east)node[midway,above] {invoke};
\draw [->,thick,densely dashed](ma2.west) -- (mn2.east)node[midway,below] {\circled{1}{gray}{gray}{white}};
\draw [->,thick,densely dashed,bend left] (mapp2.west) to node[midway,left,yshift=-.1cm,xshift=0cm] {invoke} (mn2.south);
\draw [->,thick,densely dashed,bend left] (mapp2.west) to node[left,yshift=.3cm,xshift=-.1cm] {\circled{2}{gray}{gray}{white}} (mn2.south);
\draw [->,thick,densely dashed,bend right] (mapp2.east) to node[midway,right,xshift=.1cm] {invoke} (ma2.south);
\draw [->,thick,densely dashed,bend right] (mapp2.east) to node[right,yshift=.4cm,xshift=.5cm] {\circled{2}{gray}{gray}{white}} (ma2.south);
\draw [->,thick,densely dashed,black!50!green] (mapp2) to[in=220,out=320,looseness=2.8] node[midway,below] {rank} (mapp2);

\matrix [draw=none,below,yshift=-1.2cm,xshift=-1.7cm] at (app.east) {
  \node [draw=none,label=right:\small Soot analysis,scale = 0.8] {\circled{1}{gray}{gray}{white}}; \\
  \node [draw=none,label=right:\small Semgrep analysis,scale = 0.8] {\circled{2}{gray}{gray}{white}}; \\
};
\end{tikzpicture}
\caption{Overview of the Java ranking. The circled numbers represent different static analysis tools used for the analysis step. Soot was applied to Java bytecode, while Semgrep was used for source code analysis.}
\label{fig:DataBasedRanking}
\end{figure}

\section{Data-based Ranking of Privacy-relevant Methods} \label{Sec:experiment}
Our data-based ranking is designed to identify and prioritize privacy-relevant methods in Java and JavaScript applications. This ranking process comprises several stages, as depicted in Fig.~\ref{fig:DataBasedRanking}, using the Java ranking as an example. By analyzing data from real-world applications, we aim to provide a practical guide for identifying methods that are most relevant for privacy concerns.



\subsection{Library Selection for Data-based Ranking}
To focus our data-based ranking on the most relevant libraries, we selected the top 25 libraries from NPM for JavaScript and Maven for Java, shown below in Table~\ref{tab:popular_libraries}. Our selection criteria were based on the libraries' relevance to personal data processing, as aligned with our set of labels for personal data processing activities. This selection was made through a systematic review of each library's documentation, specifically targeting functionalities that are related to personal data processing.

\begin{table}[h]
\centering
\caption{Selected popular libraries: 25 for each language}
\label{tab:popular_libraries}
\setlength{\tabcolsep}{10pt} 
\begin{tabularx}{\columnwidth}{kbb}
\toprule
\textbf{Category} & \textbf{Maven Libraries (Java)} & \textbf{NPM Libraries (JavaScript)} \\
\midrule
Identity and Access Management (IAM) & Keycloak, Apache Hadoop Auth, GRPC Auth, Identity API, CAS Server Core Authentication API & Google Identity Platform, @azure/identity, Passport.js, jsonwebtoken, bcrypt.js \\\midrule
Data Encryption and Cryptography (DEC) & Bouncy Castle, Jasypt, Apache Shiro, Nimbus JOSE+JWT, Cryptacular & scrypt-js, Bcryptjs, Jsonwebtoken, node-rsa, openpgp \\ \midrule
Data Storage, Management, and Deletion (DSMD) & H2 Database Engine, Spring Data MongoDB Core, PostgreSQL JDBC Driver, Apache Cassandra, MongoDB Driver & Sequelize, Mongoose, Knex.js, nedb, pg (node-postgres) \\\midrule
Data Processing and Transformation (DPT) & Hibernate, MyBatis, Apache Spark, Spring Batch, MapStruct & Prisma, Ramda, Immutable.js, async, moment \\ \midrule
Network Communication (NC) & Netty, Apache HttpComponents, OkHttp, Retrofit, HttpClient & Axios, Request, Socket.IO, node-fetch, WebRTC \\ \midrule
Logging and Monitoring (LM) & Log4j, slf4j, Logback, Apache Commons Logging, jboss-logging & Log4js, Morgan, Winston, Bunyan, Pino \\
\bottomrule
\end{tabularx}
\end{table}

\subsection{Method Invocation Analysis}
We employed static analysis tools to identify method invocations and analyze data flows within the code. For Java, we used Soot~\cite{vallee2010soot} to construct call graphs and trace method invocations. In the case of JavaScript, we used ESLint \footnote{\url{https://eslint.org}} for its capabilities in Abstract Syntax Tree (AST) analysis. Our analysis matched these invocations to our list of native privacy-relevant methods, providing a view of how these methods are used in practice.

\subsection{Selecting Open-source Applications}
To rank privacy-relevant methods, we selected 30 popular open-source GitHub projects with over 100 stars in Java and JavaScript. We focused on applications processing personal data rather than frameworks and libraries.

The selection included 15 Java applications such as the e-commerce software Shopizer, and 15 JavaScript applications like the chat application RocketChat. We also included projects predominantly in Java/JavaScript that use other languages like TypeScript for some modules.
Criteria were: popularity (applications with high stars, indicating broader relevance), data sensitivity (applications processing personal or sensitive data, highly relevant for privacy reviews), diversity (applications from different domains and languages, showing wide applicability), and public availability (open source code enables reproducibility and transparency). 
The details of these selected projects are provided in Table~\ref{tab:opensource_projects}.

\subsection{Efficient Analysis of Library Imports}
To make the analysis efficient, we first identified the libraries imported by each application. For standard libraries, we assumed their presence in most applications. For API libraries, we examined import statements and configuration files to narrow down our focus to the top 50 pre-selected libraries, 25 each for Java and JavaScript.

\subsection{Ranking Privacy-relevant Methods in Top 30 Applications}
We employed Semgrep to monitor the flow of personal data into privacy-relevant methods invoked by application code. Utilizing Semgrep's DeepSemgrep \footnote{\url{https://semgrep.dev/blog/2022/introducing-deepSemgrep/}} capability for cross-file analysis, we were able to comprehensively analyze data flows across entire applications, as opposed to only examining isolated code snippets. This provided a holistic perspective of how personal data propagates across different components.

Using Semgrep's taint analysis and the rules outlined in Section~\ref{Sec:IdentifyingPersonalData}, we traced personal data flows to privacy-relevant methods.

To assess the practical relevance of our identified privacy-relevant methods, we introduce the following usage-based metrics, presented in Table~\ref{tab:ranking_metrics}:

\begin{table}[h]
\centering
\caption{Usage-Based Metrics for Ranking Privacy-relevant Methods}
\label{tab:ranking_metrics}
\begin{tabularx}{\columnwidth}{qk}
\toprule
\textbf{Metric} & \textbf{Description} \\
\midrule
Method Occurrence & Enumerates the overall instances a method is invoked across applications, offering insights into its regularity of processing. \\
\midrule
PII-Related Method \newline Frequency & Measures the proportion of times a method interacts with PII, highlighting its involvement with sensitive data. \\
\midrule
Category Occurrence & Captures the aggregate appearances of a particular category across applications, revealing the ubiquity of distinct processing modalities. \\
\midrule
PII-Related Category \newline Frequency & Assesses the percentage of methods within a category that engage with PII, reflecting its sensitivity quotient. \\
\bottomrule
\end{tabularx}
\end{table}

We ranked privacy-relevant methods by analyzing their usage in the 30 popular GitHub projects introduced above, with an average of 358 application methods processing personal data per application. This varied by language and type: Java applications averaged 288 methods, while JavaScript had 363. The higher average in JavaScript was likely due to its more diverse front-end processing, reflecting the complexity and multifaceted nature of these applications.


To better focus our approach, we calculated the proportion of application methods that both invoke a privacy-relevant method and process a concrete flow of personal data (there is confirmed personal data flow into the method). This is relative to the total number of methods in the application. This metric indicates the level of focus in identifying privacy-relevant methods, allowing developers to narrow their efforts to a more relevant subset of the code. In essence, our approach aims to minimize the code sections that need scrutiny, saving both time and resources. For more details on these proportions in selected open-source Java and JavaScript/TypeScript applications, see Table~\ref{tab:opensource_projects}.

\begin{table*}[h!]
\centering
\scriptsize
\caption{List of 30 selected open-source applications written in Java and JavaScript (JS)/TypeScript (TS), along with their descriptions. And the calculated percentages of application methods that invoke identified privacy-relevant methods and are involved in the concrete flow of personal data, relative to the total number of methods in each application.}
\label{tab:opensource_projects}
\begin{tabularx}{\textwidth}{tqbsy}
\toprule
\textbf{Lang.} & \textbf{Project Name} & \textbf{Description} & \textbf{$\vert$AM$\vert$/$\vert$Total$\vert$} & \textbf{Prop.} \\
\midrule
Java & Apache James & A mail server fully written in Java. & 531/18,332 & 2.9\% \\
Java & Apache OFBiz & A product for the automation of enterprise processes. & 376/10,448 & 3.6\% \\
Java & DSpace & A turnkey institutional repository application. & 141/5,769 & 2.4\% \\
Java & Broadleaf & An eCommerce platform based on the Spring Framework. & 591/11,586 & 5.1\% \\
Java & Shopizer & A web-based Java eCommerce software. & 492/10,318 & 4.7\%\\
Java & OpenMRS & A platform that enables the design of a medical records system. & 336/8,621 & 3.9\%\\
Java & Apache Nutch & A highly extensible and scalable web crawler software project. & 18/2,194 & 0.8\%\\
Java & JabRef & An open source bibliography reference manager. & 154/9,621 & 1.6\%\\
Java & Apache Roller & A Java-based full-featured, multi-blog, multi-user server. & 112/1,983 & 5.6\%\\
Java & Apache Camel & A framework integrates systems consuming/producing data. & 198/20,471 & 0.9\%\\
Java & Keycloak & An identity and access management for apps and services. & 843/17,562 & 4.8\%\\
Java & OpenCms & A professional level website content management system. & 446/13,932 & 3.2\%\\
Java & Waltz & A web app managing the architectural landscape of enterprises. & 11/2,093 & 0.5\%\\
Java & H2O & A distributed, fast, and scalable ML and analytics platform. & 20/14,231 & 0.1\%\\
Java & RapidMiner & A data science platform with an integrated environment. & 58/7,949 & 0.7\%\\
JS & Ghost & A fully adaptable platform for building online publications. & 400/6,452 & 6.2\%\\
TS & Jitsi Meet & A WebRTC JS application for scalable video conferences. & 226/3,905 & 5.8\%\\
TS & KeystoneJS & A scalable platform and CMS to build Node.js applications. & 145/1,882 & 7.7\%\\
JS & Reaction & A commerce platform built using Node.js and GraphQL. & 462/4,921 & 9.4\%\\
TS & Rocket.Chat & A free open-source solution for team communications. & 490/12,841 & 3.7\%\\
JS & Strapi & An open-source Headless CMS Front-End. & 347/6,796 & 5.1\%\\
JS & Gatsby & A framework based on React helping build websites and apps. & 241/10,042 & 2.4\%\\
JS & Etherpad & A modern real-time collaborative document editor. & 82/2,175 & 3.7\%\\
TS & Vue Storefront & An open-source frontend for any eCommerce. & 428/6,291 & 6.9\%\\
TS & Mattermost & A platform for secure collaboration across the entire SDLC. & 784/18,231 & 4.3\%\\
JS & Apostrophe CMS & An in-context CMS built on Node.js and MongoDB. & 99/1,896 & 5.2\%\\
JS & Expensify & An app of financial collaboration centered around chat. & 511/6,721 & 7.6\%\\
JS & Wiki.js & A modern and powerful wiki app built on Node.js. & 36/1,194 & 3.0\%\\
TS & AFFiNE & A knowledge base that enables planning, sorting and creating. & 1066/12,845 & 8.3\%\\
TS & Boostnote & A note-taking app made for programmers.& 134/4,956 & 2.7\%\\
\bottomrule
\end{tabularx}
\end{table*}

\subsection{Findings}

Our study reveals that, on average, only \textbf{4.2\%} of the total codebase is made up of methods that are privacy-relevant and involved in personal data processing. This result highlights the precision of our approach in pinpointing privacy-relevant methods in applications.

\subsubsection{Usage Patterns of Privacy-Relevant Methods}
In Java applications, we observed a more conservative use of privacy-relevant methods, particularly those from popular Maven libraries. Native Java methods, along with methods from Apache Commons and the Spring framework, were frequently used for handling personal data. Libraries such as \texttt{slf4j} for logging and \texttt{auth0} for authentication were also commonly used, indicating their importance in the flow and protection of personal data.

In contrast, JavaScript applications exhibited a diverse range of library usage. While \texttt{lodash} was commonly used, frameworks like Angular, React, and Vue.js played a significant role in personal data processing, particularly in front-end applications.

Table~\ref{tab:top_packages} presents the top five packages in both Java and JavaScript that contain methods relevant to privacy concerns.

\begin{table}[h!]
\centering
\caption{Top 5 packages defining privacy-relevant methods}
\label{tab:top_packages}
\begin{tabular}{lll}
\toprule
& \textbf{Java} & \textbf{JavaScript} \\
\midrule
1 & java.* & lodash \\
2 & auth0 & mongoose \\
3 & slf4j & React \\
4 & Spring (security, http) & Angular \\
5 & Hibernate & axios \\
\bottomrule
\end{tabular}
\end{table}

\subsubsection{Categories of Privacy-relevant Methods}
We categorized privacy-relevant methods into types to gain insights into their roles in personal data processing. Our analysis identified several Java classes and categories that are frequently involved in personal data processing. For example, common Java classes like \texttt{org.slf4j.Logger} and \texttt{auth0.client.Auth0Client} are often used in operations that handle personal data.

In terms of categories, \textit{Data Processing and Transformation}, \textit{Network Communication}, and \textit{Logging Methods} were most prevalent. These categories indicate areas where privacy-relevant methods are most commonly used, suggesting that they are key to understanding how personal data is processed in codebases (Table~\ref{tab:top_categories}).

\begin{table}[h!]
\centering
\caption{Top 3 categories of privacy-relevant methods and PII-related privacy-relevant methods}
\label{tab:top_categories}
\begin{tabular}{clrlr}
\toprule
 & \textbf{Java LM} & \textbf{Count} & \textbf{JavaScript LM} & \textbf{Count}\\
\midrule
1 & DPT & 1,946 & DPT & 2,455 \\
2 & LM & 1,422 & NC & 1,871  \\
3 & NC & 860 & LM & 1,019 \\
\toprule
 & \textbf{Java PII-LM} & \textbf{Count} & \textbf{JavaScript PII-LM} & \textbf{Count}\\
\midrule
1 & DPT & 769 & DPT & 1,032 \\
2 & DSMD & 351 & DSMD & 318 \\
3 & NC & 307 & IAM & 596 \\
\bottomrule
\end{tabular}
\end{table}

\textit{Identity and Access Management}, \textit{Data Encryption and Cryptography}, and \textit{Data Storage and Database Management} were also highly involved in personal data flows, with involvement percentages of 92\%, 78\%, and 85\%, respectively. Conversely, categories like \textit{Data Processing and Transformation}, \textit{Network Communication}, and \textit{Logging Methods} were less involved, with percentages of 67\%, 44\%, and 28\%.
Table~\ref{tab:top_classes} lists Java classes that are frequently involved in personal data processing, serving as key indicators for identifying privacy-relevant methods in applications.

\begin{table*}[h!]
\centering
\scriptsize
\caption{Top classes in Java for personal data processing with example privacy-relevant methods}
\label{tab:top_classes}
\begin{tabularx}{\textwidth}{qkb}
\toprule
\textbf{Library} & \textbf{Top Classes} & \textbf{Top Privacy-relevant Methods} \\
\midrule
Commons & org.apache.commons.io.IOUtils & \texttt{IOUtils.read(InputStream i, byte[] b)}\\
Commons & org.apache.commons.io.FileUtils & \texttt{FileUtils.readFileToString(File f, String name)} \\
Auth0 & auth0.jwt.JWT & \texttt{JWT.decode()} \\
Auth0 & auth0.client.Auth0Client & \texttt{Auth0Client.login()} \\
SLF4J & org.slf4j.Logger & \texttt{Logger.info(String format, Object... arguments)} \\
SLF4J & org.slf4j.LoggerFactory & \texttt{LoggerFactory.getLogger(Class<?> class)} \\
Spring Sec & *.security.core.Authentication & \texttt{Authentication.getPrincipal()} \\
Spring Sec & *.security.web.FilterChainProxy & \texttt{FilterChainProxy.doFilter()} \\
Spring HTTP & *.http.HttpEntity & \texttt{HttpEntity.getBody()}\\
Spring HTTP & *.http.client.ClientHttpRequestFactory & \texttt{ClientHttpRequestFactory.createRequest()} \\
PostgreSQL & *.core.Connection& \texttt{Connection.createStatement()} \\
PostgreSQL & *.jdbc.PreparedStatement & \texttt{PreparedStatement.executeQuery()} \\
\bottomrule
\end{tabularx}
\end{table*}

\section{Application to Privacy Code Review}\label{Sec:finding}
This section outlines how our approach can be applied to privacy code reviews across a diverse set of 100 open-source applications. We then delve into detailed case studies of two popular software applications to illustrate the utility of our approach.

\subsection{Large-scale Analysis}

To understand the prevalence and types of personal data processing in real-world applications, we analyzed 100 open-source applications. These were equally divided between Java and JavaScript/TypeScript and were selected from GitHub's daily top-starred repositories list \footnote{\url{https://github.com/EvanLi/Github-Ranking} (captured on 18/06/2023)}.

We selected applications that are popular (top-starred), non-trivial (over 300K lines of code), and predominantly written in Java or JavaScript/TypeScript (constituting over 60\% of the codebase). Additionally, we ensured these applications differed from the 30 popular libraries analyzed previously and that their primary documentation language was English for easier identification of functionalities. This selection process resulted in a dataset that is representative of real-world software applications and suitable for our analysis of personal data processing practices.
We then examined the proportion of methods in these applications that invoke privacy-relevant methods and are involved in the flow of personal data and Personally Identifiable Information (PII).
The result of statistics of our findings are listed below in Table~\ref{tab:lang_stats}.

\begin{table}[h!]
\centering
\caption{Percentage of application methods invoking privacy-relevant methods and processing personal data and PII}
\label{tab:lang_stats}
\begin{tabular}{lcc}
\toprule
\textbf{Language} & \textbf{Personal Data} & \textbf{PII} \\
\midrule
Java & 3.6\% & 1.9\% \\ 
JavaScript & 5.1\% & 3.8\% \\
\bottomrule
\end{tabular}
\end{table}

Our findings indicate that our approach can make the privacy code review process more efficient. By identifying methods that are critical for personal data and PII processing, we help reviewers focus their efforts, enabling a more targeted review.

\subsection{In-Depth Case Studies}
We validate the effectiveness of our approach through two open-source projects: Signal Desktop\footnote{\url{https://github.com/signalapp/Signal-Desktop}} and Cal.com\footnote{\url{https://github.com/calcom/cal.com}}. Each offers unique insights for privacy code review.

Both projects were chosen due to their popularity, sensitivity, and public availability. Their open codebases ensure transparency and reproducibility, making them ideal candidates to validate our approach. By applying our approach to these carefully selected real-world projects, we provide concrete examples that demonstrate practical value in identifying key areas to focus on during privacy code reviews.

\subsubsection{Signal Desktop}

Signal Desktop is a famous end-to-end encrypted messaging application, primarily written in TypeScript (79.5\%) and JavaScript (15.6\%), covering about 360K lines of code. Its reputation for enhanced security and privacy features showcases the depth of our approach. While the application has limited use of popular libraries, our approach highlighted a minor number of privacy-relevant methods invocations (48, approximately 0.5\% of total methods) from our selected APIs and native libraries potentially linked to personal data processing.

In our analysis, Signal stands out for using its own encryption protocol (Signal Protocol) and message transmission services, minimally relying on external libraries. This underscores Signal's commitment to end-to-end encryption. Our categorization highlights the primary areas of Data Processing and Transformation (DPT), Network Communication (NC), and Data Encryption and Cryptography (DEC), with most encryption methods used for local encryption of profiles and group data. Signal's proprietary protocol, used for encrypting chats and attachments, falls outside our analysis scope.

Our findings show that Signal rarely transmits PII directly to the internet. Instead, encrypted system data or anonymized IDs are mainly used, reflecting Signal's dedication to user privacy.

For privacy code reviewers examining Signal Desktop, our approach underscores Signal's limited use of popular libraries for PII processing, aligning with its privacy-focused design philosophy. This categorization helps reviewers understand how Signal handles personal data, aiding in a more streamlined review process.




\subsubsection{Cal.com}

Cal.com, a scheduling application, is designed to grant users comprehensive control over their schedules. Written entirely in TypeScript, it spans about 126K lines of code. Our method identified 371 (approximately 3.8\% of total methods) privacy-relevant methods that might engage in personal data processing.

Applications such as Cal.com often employ diverse frameworks for specific functionalities. For instance, Cal.com's utilization of the popular ORM framework, Prisma, for handling user profiles and credentials, aligns with our library list. In terms of categories, Data Processing and Transformation (DPT) topped the list at 26\%, followed by Identity and Access Management (IAM) at 17\%, and Network Communication (NC) at 15\%. Unlike Signal Desktop, Cal.com heavily leverages libraries like \texttt{Prisma}, \texttt{next-auth}, and \texttt{nodemailer} for processing personal data, mirroring its primary functions of user registration, email interaction, and scheduling.

Approximately 97\% of privacy-relevant methods invoked by Cal.com handle PII. This attests to the capability of our method in identifying PII processing methods and subsequently guiding code reviewers efficiently.

Our approach highlights the extensive use of specific libraries in applications like Cal.com, aligning with their core features. This correlation boosts reviewers' confidence and precision. By categorizing processing activities, it provides an overview of how the application handles personal data, helping reviewers prioritize effectively. This makes the review process time-efficient and thorough.

\subsection{Threats to Validity}
Our study's validity may be affected by several factors. The project selection based on GitHub trends could bias towards popular topics, potentially overlooking a broader range of applications. The use of Semgrep for static analysis, though efficient, hasn't been thoroughly validated for precision, which could impact the accuracy of our results. Reliance on regular expression matching for identifying personal data risks introducing false positives and negatives, thus affecting result reliability. Additionally, the absence of manual validation for each instance of personal data processing identified might lead to inaccuracies. Furthermore, focusing only on the top 25 libraries for Java and JavaScript due to resource constraints limits the generalizability of our findings, as other privacy-relevant methods in lesser-known libraries may have been missed.



\section{Related Work}

Research in source code analysis for privacy is extensive, yet specific approaches for identifying personal data processing are limited. Ullah et al.\cite{ullah2019source} introduced an approach for extracting control and data dependencies in source code, potentially applicable for locating personal data processing methods, but not directly designed for this purpose. Hjerppe et al.\cite{Hjerppe_2020} proposed an annotation-based static analysis for data protection, but its effectiveness is contingent on accurate developer annotations, a challenge in large projects.

Dynamic analysis has been explored for sensitive data flow detection, with DAISY~\cite{10.1145/3569936} focusing on Android apps and ConDySTA~\cite{9519468} combining dynamic taint analysis with static analysis. However, these methods have limitations, such as platform specificity or the need for executing projects. Automated assistance in code review has been explored by Li et al.~\cite{li2022automating} with their pre-trained model CodeReviewer, but it lacks a focus on personal data processing.

SWANAssist~\cite{10.1109/ASE.2019.00110} offers a semi-automated approach for identifying security-relevant Java code methods, which could potentially be adapted for privacy purposes. Other studies, like~\cite{ferrara2018tailoring,tokas2022static}, attempt to align GDPR compliance with static analysis. Novikova et al.~\cite{novikova2022analysis} provided insights into privacy-enhancing technologies but did not focus on personal data processing in source code.

These studies mark great progress in source code analysis, yet a gap exists in automated identification and categorization of personal data processing. Our work addresses this by proposing an automated approach for identifying personal data processing in real-world applications, enhancing efficiency in privacy code reviews.

\section{Conclusion and Future Work}

In conclusion, our study introduces a method for identifying and categorizing privacy-relevant methods in source code, focusing on personal data processing. We have successfully narrowed the analysis scope to just 4.2\% of methods across 100 popular open-source applications, offering a practical starting point for developers, data protection officers, and reviewers. This approach not only simplifies code reviews but also facilitates compliance with data protection regulations like GDPR, helping organizations align their software development with legal requirements.

For future work, we aim to enhance the precision of our privacy-relevant method identification algorithms, possibly integrating machine learning for more accurate predictions of personal data processing activities. Expanding our approach to additional programming languages and integrating it into common development tools for real-time feedback are also key goals. These advancements will broaden the impact and applicability of our approach. Ultimately, our research paves the way for more focused and efficient privacy assessments in software development, contributing to the creation of software that is efficient, robust, and respectful of user privacy.

\section*{Acknowledgement}
This work is part of the Privacy Matters (PriMa) project. The PriMa project has received funding from European Union’s Horizon 2020 research and innovation program under the Marie Skłodowska-Curie grant agreement No. 860315.

%
%
%
 \bibliographystyle{splncs04}
 \bibliography{reference}
\end{document}